# Motions of Classical Charged Tachyons

by
Mark Davidson
Spectel Research Corporation
807 Rorke Way
Palo Alto, CA 94303
email: mdavid@spectelresearch.com

## Abstract

It is shown by numerical simulation that classical charged tachyons have self-orbiting helical solutions in a narrow neighborhood of certain discrete values for the velocity when the electromagnetic interaction is described by Feynman-Wheeler electrodynamics. The force rapidly oscillates between attractive and repulsive as a function of velocity in this neighborhood. Causal electrodynamics is also considered, and in this case it is found that when the force is attractive the tachyon loses energy to radiation. Only certain narrow ranges of velocity give attractive forces, and a geometrical derivation of these special velocities is given. Possible implications of these results for hidden variable theories of quantum mechanics are conjectured.

### Résumé

À l'aide de la simulation numérique on démontre que les tachyons chargées classiques ont des solutions hélicoïdales et auto-orbitantes dans une étroite région de valeurs distinctes pour la vitesse quand on exprime l'interaction électromagnétique avec l'électrodynamique Feynman-Wheeler. La force oscille rapidement entre l'attraction et le recul comme fonction de la vitesse dans cette région. Aussi l'électrodynamique causale est considérée, et dans ce cas on constate que le tachyon perd de l'énergie à la radiation avec la force attractive. Seules quelques étendues étroites de vitesse remettent les forces attractives, et on donne une dérivation géométrique de ces vitesses exceptionnelles. Des significations possibles pour ces résultats sont présentées par conjecture pour les théories de variables dérobées de la mécanique quantique.

## PACS Numbers: 41.20.-q  14.80.-j

# 1. Introduction

Classical tachyons have been extensively studied in the physics literature [1-7]. The peak period of interest was in the 1960s and 70s. A cloud of questions arose over issues of causality [8], and as tachyons had not been observed in nature, interest in them waned. Arguments were made to counter the causality objections [9], and the issue remains in dispute. There are several reasons why tachyons are still of interest today, and in fact interest may be increasing. First, many string theories have tachyons occurring as some of the particles in the theory [10], although they are generally regarded as unphysical in those theories. There are also several recent papers that assert experimental evidence that some neutrinos are tachyons [11,12]. There is a new and extensive re-analysis of tachyon dynamics [13]. There is much discussion in the physics literature in recent years of superluminal connections implied by quantum mechanics and by the evanescent wave phenomenon of light optics as well as quantum tunneling. These and other recent developments show that tachyons are still a timely subject for investigation.

Tachyon trajectories can conceivably intersect their own past and future light cones [4], and if they were charged they would experience an electromagnetic self force that has no analog for particles moving slower than light. If a tachyon were moving very fast, then it could intersect with its own light cone at a large number of points in its own past or future. The resulting theory looks to be extremely complex and fertile for investigation, having more in common with a many body problem than with a classical single particle trajectory problem. According the the analysis of Ey and Hurst [5], charged tachyons will not experience a local self force (the type that occurs in the Lorentz Dirac Equation), and they concluded from this that consequently a tachyon will not radiate. Tachyons will experience a self force when they cross their own light cone as we consider here, and in this case they can radiate at least in the usual causal electrodynamics with retarded potentials. In the Feynman-Wheeler formulation of electrodynamics the Tachyons won't radiate.

Recami [6] has presented a thorough analysis of the various possibilities for classical equations of motion for tachyons. We show that in Feynman-Wheeler action at a distance electrodynamics these equations have many closed circular or helical orbits, depending on the speed of the tachyon in the circular orbit. Some speeds give bound orbits while others do not. The self-orbiting tachyon thus appears to be moving slower than the speed of light on average, and to have an intrinsic angular momentum. One might intuitively expect that a tachyon which is moving in a circle so fast as to see its own image – ie. to intersect its own light cone – would be repelled from its own image because like charges repel. This is not the case because magnetic and relativistic effects must be included which lead to net attraction in some cases, depending on the speed of the orbit.

We first explore the Feynman-Wheeler form for Electrodynamics and show that helical solutions exist. Then we consider causal electrodynamics and calculate the azimuthal force which causes the motion to be non-helical. The radial force for helical motion in this case is the same as for Feynman-Wheeler interaction. The calculations are too complicated to perform analytically, and so numerical computations are presented.

## 2. Tachyon Equations of Motion

Tachyon classical dynamics are not unique, but rather there are different possibilities which are described for example by Recami [6]. The most commonly accepted form for the Tachyon equations of motion are (Recami[6], in units where c = 1):

$$\mathbf{F} = \frac{d}{dt}(\frac{m_0 \mathbf{V}}{\sqrt{\mathbf{V}^2 - 1}}) \tag{1}$$

Recami argues that the force equation could with equal plausibility be chosen to be (see section 6'14 in Reference 6)

$$\mathbf{F} = -\frac{d}{dt}(\frac{m_0 \mathbf{V}}{\sqrt{\mathbf{V}^2 - 1}}) \tag{2}$$

where in both of these equations the mass $m_0$ is positive and real. The two equations differ by an overall sign. These equations are completely different and have very different solutions. Both of them have circularly orbiting tachyon solutions in Feynman Wheeler electrodynamics.

We shall first illustrate the solution for the time symmetric action at a distance electrodynamics of Fokker, Feynman, and Wheeler [14-15]. The slower than light problem of a relativistic two-body bound state has been solved for this type of interaction [16] which lends encouragment toward looking for a solution in this context in the present case.

## 3. Action-at-a-Distance Electrodynamics

When the force is due solely to electromagnetic effects, we may write for equation 1 of motion of the electrically charged tachyon:

$$\frac{d}{dt}(\frac{m_0}{\sqrt{\mathbf{v}^2/c^2 - 1}}\frac{d\mathbf{x}}{dt}) = q[\mathbf{E} + \frac{\mathbf{v}}{c} \times \mathbf{B}] \tag{3}$$

where $q$ is the charge of the tachyon and

$$\mathbf{E} = -\nabla \Phi - \frac{1}{c}\frac{d\mathbf{A}}{dt}; \quad A^m = (\mathbf{A}, i\Phi) \tag{4}$$

and where

$$A^m = \frac{1}{c} \iint \frac{J^m(\mathbf{x}',t')}{R} [\frac{1}{2}\boldsymbol{d}(t'+\frac{R}{c}-t) + \frac{1}{2}\boldsymbol{d}(t'-\frac{R}{c}-t)]d^3x'dt', R = |\mathbf{x}-\mathbf{x}'| \quad (5)$$

where the currents $\boldsymbol{J}$ are due to all the relevant charges. The expression (5) is symmetrical between the future and the past since the advanced and retarded conditions enter with equal weights in the action-at-a-distance approach..

In (5) must now be included the self interaction possibilities since the tachyon moves faster than light and it can therefore interact through the retarded and advanced potential with its own trajectory. Thus in (5), $\boldsymbol{J}$ includes not only the current density produced by all the other relevant particles, but also that of the particular particle whose motion the equation describes. Excluded from $\boldsymbol{J}$, however, is the singular contribution arising from the particle's present location, just as this is also excluded in the slower than light case. Since tachyon's can intersect their own light cone, but slower than light particles (tardyons or bradyons) cannot, the tachyon case is much richer in solutions as has been pointed out by Cawley [4].

Next we consider an isolated electrically charged tachyon. The only contributions to $\boldsymbol{J}$ in equation 5 is then from the particle's own trajectory intersecting its own light cone.

For any particle, tachyon or tardyon, which moves strictly forward in time we may write the contribution to $\boldsymbol{J}$ as

$$J^m(\mathbf{x},t) = qc\boldsymbol{b}^m\boldsymbol{d}^3(\mathbf{x}-\mathbf{r}(t)), \quad \boldsymbol{b}^m = (\frac{1}{c}\frac{d\mathbf{x}}{dt},i) = (\boldsymbol{\beta},i) \quad (6)$$

The advanced and retarded potentials can be written

$$A^m_{\substack{\text{Ret} \\ Adv}}(\mathbf{x},t) = q\int \frac{\boldsymbol{b}^m(t')}{|\mathbf{x}-\mathbf{x}(t')|}\boldsymbol{d}(t'\pm\frac{|\mathbf{x}-\mathbf{x}(t')|}{c}-t)dt' \quad (7)$$

Let us use the notation

$$\mathbf{R}(\mathbf{x},t') = \mathbf{x}-\mathbf{x}(t'); \quad R(\mathbf{x},t') = |\mathbf{R}(\mathbf{x},t')| \quad (8)$$

So that $\mathbf{R}$ is the vector pointing from the source point on the particle trajectory to a test point $\mathbf{x}$. Then defining

$$K_{\substack{\text{Re}\,t\\Adv}}(\mathbf{x},t')=\frac{d}{dt'}(t'\pm\frac{R(\mathbf{x},t')}{c})=1\mp\hat{\mathbf{n}}\cdot\boldsymbol{\beta}\quad\text{where }\hat{\mathbf{n}}=\frac{\mathbf{R}(x,t')}{R(x,t')}\tag{9}$$

we can write

$$\Phi^{\substack{\text{Re}\,t\\Adv}}=e\left[\frac{1}{KR}\right]_{\substack{ret\\adv}}\quad;\quad\mathbf{A}^{\substack{\text{Re}\,t\\Adv}}=e\left[\frac{\boldsymbol{\beta}}{KR}\right]_{\substack{ret\\adv}}\tag{10}$$

where $\mathbf{b}$, K and R are evaluated at the retarded or advanced time as determined by the case. The gradient operator acting on any of these fields can be expressed in terms of the R derivative

$$\nabla\Rightarrow(\nabla R)\frac{\partial}{\partial R}=\hat{\mathbf{n}}\frac{\partial}{\partial R}\tag{11}$$

Then we find the following expressions for electric and magnetic fields analogous to the slower than light case:

$$\mathbf{E}^{\substack{ret\\adv}}=\left[q\frac{(\hat{\mathbf{n}}\mp\boldsymbol{\beta})}{K^3R^2}(1-\boldsymbol{\beta}^2)+\frac{q}{cK^3R}\hat{\mathbf{n}}\times\left((\hat{\mathbf{n}}\mp\boldsymbol{\beta})\times\dot{\boldsymbol{\beta}}\right)\right]_{\substack{ret\\adv}}\tag{12}$$

and

$$\mathbf{B}^{\substack{ret\\adv}}=\pm\hat{\mathbf{n}}_{\substack{ret\\adv}}\times\mathbf{E}^{\substack{ret\\adv}}\tag{13}$$

Notice an interesting feature of equation 12. K as defined in equation 9 can be zero. Whenever K vanishes, the electric field in (12) can be singular even if R is not zero. This singular behavior has no analogue in slower than light electrodynamics. The vanishing of K defines the Cerenkov cone, and singularities on this cone were also noted in [5].

## 4. Circular Motion Solutions

We now look for solutions to the equations of motion (1) in the form of circular motion:

$$x(t)=r\cos(?\,t);\quad y(t)=r\sin(?\,t);\quad z=0\tag{14}$$

and where the speed is superluminal

$$v = r? > c \qquad (15)$$

The helical trajectory may instersect the lightcone of a testpoint on the trajectory at a number points as is illustrated in Figure 1. The number of intersections is a function only of the particle's speed. We must sum the force contribution from all of these intersection points. The points come in symmetrical pairs, each retarded point being the reflection of the correspondng advanced point.

Depending on how fast the tachyon is moving, the trajectory will intersect the future and past light cones symmetrically at least at one point. Consider Figure 2 as we calculate the fields at an arbitrary test point along the trajetory which are due to a pair of light cone intersections. The total field at the test point is given by:

$$\mathbf{E} = \frac{1}{2}\left(\mathbf{E}_{ret} + \mathbf{E}_{adv}\right); \quad \mathbf{B} = \frac{1}{2}\left(\mathbf{B}_{ret} + \mathbf{B}_{adv}\right) \qquad (16)$$

For circular motion $\dot{\mathbf{B}}$ points in the negative radial direction in Figure 2. Notice from figure 2 and equation 9 that for this case of reflection pairs of intersecting points that

$$K_{ret} = K_{adv} \equiv K \qquad (17)$$

Neither the retarded nor the advanced force is in the radial direction. However, it is straightforward to show that when one adds the contribution of a pair of advanced and retarded symmetrical intersection points, that the force acting on the test point due to this contribution becomes a radial force. Schild [16] noted this same fact for slower than light circular orbiting particles.

It follows that R is proportional to the radius r if $\beta$ is fixed and also that $\left|\dot{\beta}\right| = c\beta^2 / r$.

Therefore from (12) it follows that the radial force acting on the test point has the overall form :

$$F_r = -\frac{q^2}{r^2} Z(\beta) \qquad (18)$$

where Z is a dimensionless function depending only on $\beta$ and where positive values of Z correspond to attraction to the center of the helix and negative values repulsion. A plot of Z is presented in Figure 3 up to $\beta = 20$. Whenever Z is positive a helical motion solution exists to (1). When Z is negative, a helical solution exists for equation (2).

The equations of motion become simply:

$$\frac{m_0}{\sqrt{\mathbf{v}^2/c^2-1}}\frac{\mathbf{v}^2}{r}=\frac{q^2}{r^2}Z(\ss) \Rightarrow r=\frac{\sqrt{\ss^2-1}}{m_0 c^2 \ss^2}q^2 Z(\ss) \qquad (19)$$

Notice that for certain values of β that Z becomes discontinuous in Figure 3. These discontinuities occur when the Cerenkov cone from the source point intersects the test point. The next section explains this behavior.

## 5. Geometrical Explanation of the Discontinuities in Z which occur at Special Values of β

The numerical simulations show singular discontinuous behavior in the radial and azimuthal forces when the velocity has certain discrete values. The occurrence of these singularities can be understood by the following simple geometrical analysis. Consider the retarded fields produced by the circular motion of (14). The retardation condition which determines the times t' (sourcepoints) on the trajectory that can intersect with the position of the particle at time t (testpoint) is

$$|x(t)-x(t')|=c(t-t'), \quad \text{where } t > t' \qquad (20)$$

The reader should note that (20) is not true for arbitrary pairs of points on the helix, but only for pairs of points which are connected by a null vector. Owing to the circular form of the motion, this condition may be written as

$$r^2\left|e^{i\gamma(t-t')}-1\right|^2=c^2(t-t')^2, \, t>t' \qquad (21)$$

further defining
$$\ss = ? \, r/c \quad \text{and} \quad t = c(t-t')/r \qquad (22)$$

the null condition becomes

$$2-2\cos(\ss t)=t^2, \, t > 0 \qquad (23)$$

The larger that β is the more solutions there are to this equation. Defining:

$$f(t,\ss)=2-2\cos(\ss t)-t^2, \quad t > 0 \qquad (24)$$

Then the null points are determined by the equation

$$f(t,\ss)=0, \, t > 0 \qquad (25)$$

Noticing that the following properties are true:

$$f(0, ß) = 0 \qquad (26)$$

$$f(t, ß) \approx t^2(ß^2 - 1), \text{ for small } t \qquad (27)$$

$$f(\infty) < 0 \qquad (28)$$

It follows from these results that for $\beta > 1$ there is at least one solution for $\tau$ which satisfies the null condition. Now define:

$$N(ß) = \text{The number of solutions to } f = 0 \text{ for positive } t \qquad (29)$$

This function starts off at 1 and increases discontinuously by jumps of two at every point $\beta$ at which $f$ has a solution to the following two simultaneous equations:

$$f(t, ß) = 0; \quad \frac{\partial f(t, ß)}{\partial t} = 0 \qquad (30)$$

These are equivalent to:

$$\varphi = ßt \qquad (31)$$

$$2 - 2\cos(\varphi) - \varphi \sin(\varphi) = 0 \qquad (32)$$

$$ß = \sqrt{\boldsymbol{j} / \sin(\boldsymbol{j})} \qquad (33)$$

Equation 32 is a single transcendental which can be solved easily by numerical means. Table I shows the first 15 values of $\beta$:

| | | |
|---|---|---|
| 4.603338848751701 | 20.39583252184294 | 36.11446976533017 |
| 7.789705767492714 | 23.54070189773618 | 39.25717095448966 |
| 10.94987986982622 | 26.68479810180271 | 42.39970774262564 |
| 14.10169533046915 | 29.82836607105987 | 45.54211418676631 |
| 17.24976556755881 | 32.97155711433862 | 48.68441554248154 |

Table I Singular $\beta$ values

Next it will be proven that at $K_{\text{ret}}$ vanishes at these singular values where the number of roots to the null equation changes discontinuously by increments of 2. This also causes the Force to be singular in a neighborhood of these points. The velocity of the source-point is given by:

$$\text{ß}_x(t') = -r? \, \sin(? \, t'); \quad \text{ß}_y = r? \, \cos(? \, t') \tag{34}$$

and $K_{\text{ret}}$ from (9) is given by

$$K_{ret}(t,t') = 1 - \text{ß}(t') \cdot (\mathbf{x}(t) - \mathbf{x}(t')) / \left| \mathbf{x}(t) - \mathbf{x}(t') \right| \tag{35}$$

which can be rewritten as

$$K_{ret}(t,t') = 1 - \frac{r?}{c} \, \text{Im}(e^{iw(t-t')} - 1) / \left| e^{iwt} - e^{iwt'} \right| \tag{36}$$

or as

$$K_{ret}(t,t') = 1 - \boldsymbol{b} \, \sin(\boldsymbol{j}) / \sqrt{2 - 2\cos(\boldsymbol{j})}$$

but at a singular null point we have also from above that:

$$2 - 2\cos(\boldsymbol{j}) - \boldsymbol{j} \, \sin(\boldsymbol{j}) = 0 \tag{37}$$

and therefore

$$K_{ret}(t,t') = 1 - \boldsymbol{b} \, \sqrt{\sin(\boldsymbol{j})} / \sqrt{\boldsymbol{j}} \tag{38}$$

but also at a singular null solution we have from above

$$\text{ß} = \sqrt{\boldsymbol{j} / \sin(\boldsymbol{j})} \tag{39}$$

and therefore $K_{\text{ret}} = 0$ at a singular null solution.

Therefore, at the singular values of β, two roots to the null condition merge into one as the singularity is approached from above. Below the singularity, the roots disappear altogether. As the singularity is approached, the two $K_{\text{ret}}$ values both approach zero but with opposite sign, and the forces due to each of these sourcepoints on the testpoint approaches infinity, but again with opposite sign. The net force on the testpoint is thus the difference between two large numbers, and numerical results suggest that the behavior of the net force in a neighborhood of a singularity is a very complicated function of the velocity β showing extremely complex behavior and changing sign many times in a very small interval. Three different computer programs were used to analyze this behavior. The first two used 64 bit double precision variables, but it was found that this was inadequate precision for this problem, although they did show complex behavior. The third program used the extended precision capabilities of Mathematica, and the results obtained were stable to further increases in the working precision of the program. The results presented in the singular neighborhood were done with a working precision of 300 decimal places.

## 6. Summary of Numerical Results

The numerical calculations take a tachyon which is constrained to move on an exact circle with constant speed, and they calculate the radiative self force on the particle. Both Feynman-Wheeler elecotrodynamics and causal electrodynamics give the same radial force, but for Feynman-Wheeler electrodynamics the azimuthal force is zero.

The first general feature that one can see from Figure 3 is that the radial force appears to be repulsive for all velocities. There are discontinuities in the repulsive force, and these occur at the discrete velocities of the previous section (see Table I).

In the neighborhood of the discontinuities, the repulsive force is a very complex function of velocity as is illustrated in Figure 4 for the first singular velocity in Table I. The sign of the force in this neighborhood can be either repulsive or attractive, and the magnitude can be very large. Finer and finer sampling of the force shows even more structure with the variability of the force with velocity growing more violent as the singular velocity value is approached from above. This behavior may be understood using the results of section 5 which showed that the singularities occur when two source points merge together. When the two source points are just slightly separated, they subtract from one another in the resultant field expressions so that the resulting force on the tachyon is a difference between two large numbers leading to the large variability of the force with velocity. Notice the narrow range of velocities in Figure 4

The calculations were done with a numerical precision of 300 decimal places using Mathematica, and the results were tested for stability as the working precision was increased and they were stable. The calculation was originally coded using standard 64 bit floating point variables, but this level of precision was insufficient to do the calculation since there is a subtraction of two large numbers resulting in the net force. This necessitated a switch to Mathematica. Thus it can be concluded that for special discrete velocities the tachyon experiences an attractive force (positive values of Z), but this happens only in a neighborhood of the singular velocity values. In this neighborhood the force oscillates rapidly between attractive and repulsive as illustrated in Figure 4.

For causal electrodynamics the azimuthal force is numerically found to be nonzero. The ratio of the azimuthal to the radial force is positive for all the points calculated as illustrated in Figures 5 and 6. When the radial force is attractive, the azimuthal force is opposite in direction to the particle's velocity, and so it is tending to reduce the energy of the particle and can be interpreted as energy lost to radiation. But when the radial force is repulsive, the azimuthal force is in the same direction as the velocity and so it is tending to increase the kinetic energy of the particle. If we imagine a situation where only the radius of the orbit is constrained but the velocity is free to vary, we might get some insight into this situation.  Due to the azimuthal force the tachyon in this case would gain energy, and it would consequently slow down until it reached one of the singular velocities, and at that point the force would fluctuate between positive and negative values possibly trapping the particle's velocity at the singular value.

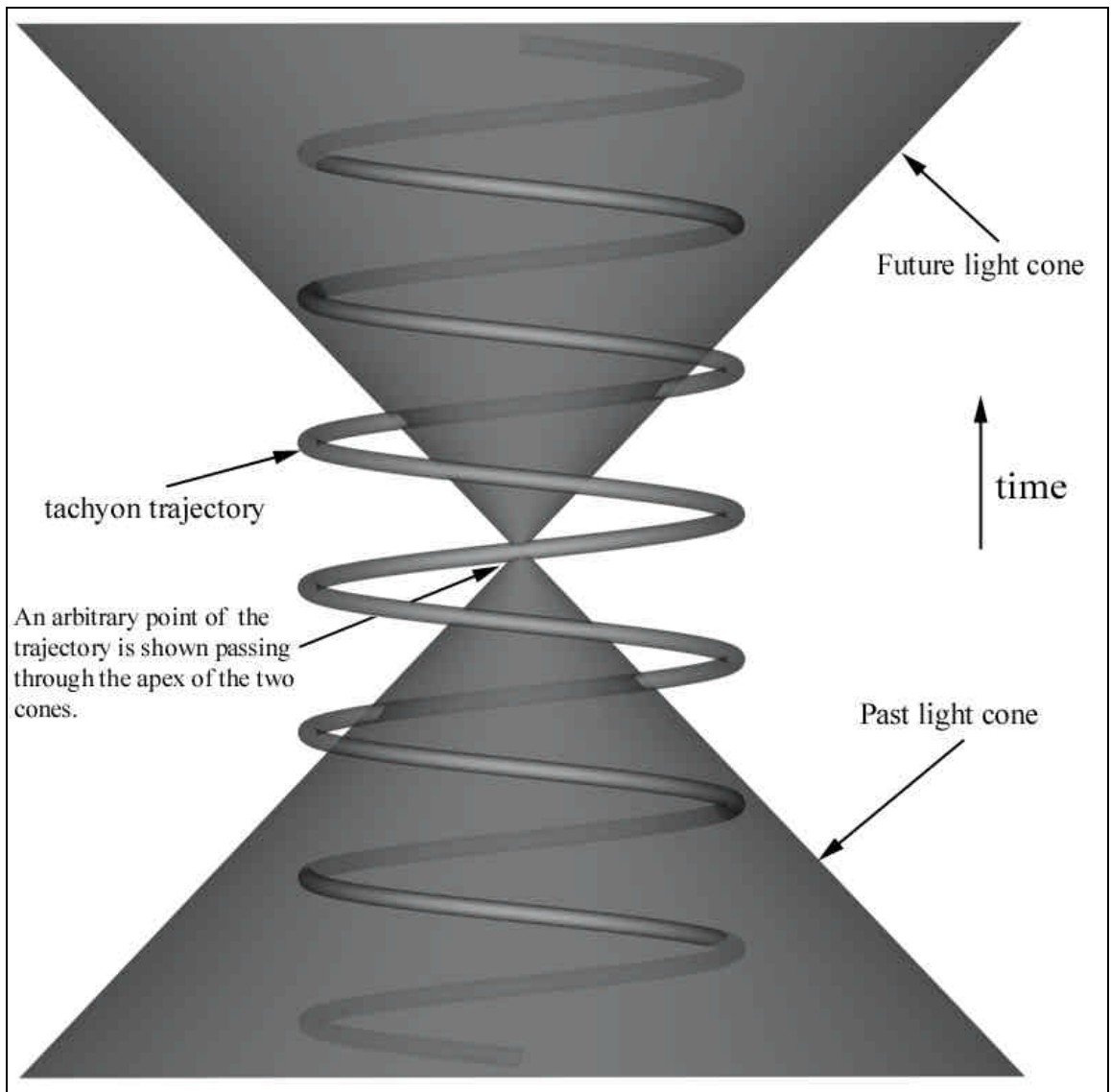

tachyon trajectory

An arbitrary point of the trajectory is shown passing through the apex of the two cones.

Future light cone

time

Past light cone

**Figure 1**  A tachyon moving in a helix and intersecting its own light cone

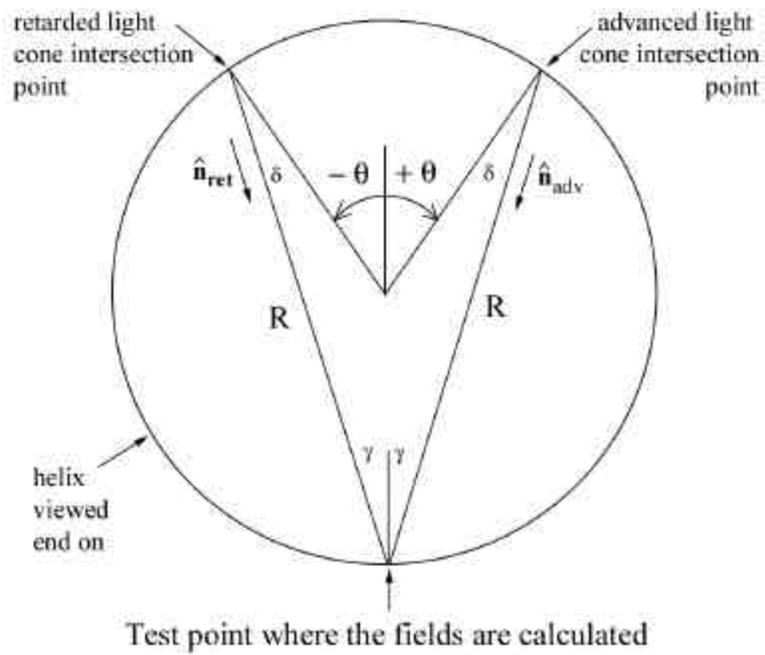

**Figure 2** Light cone intersection points come in symmetrical pairs, one point the intersection with the retarded light cone and the other the intersection point of the advanced light cone.

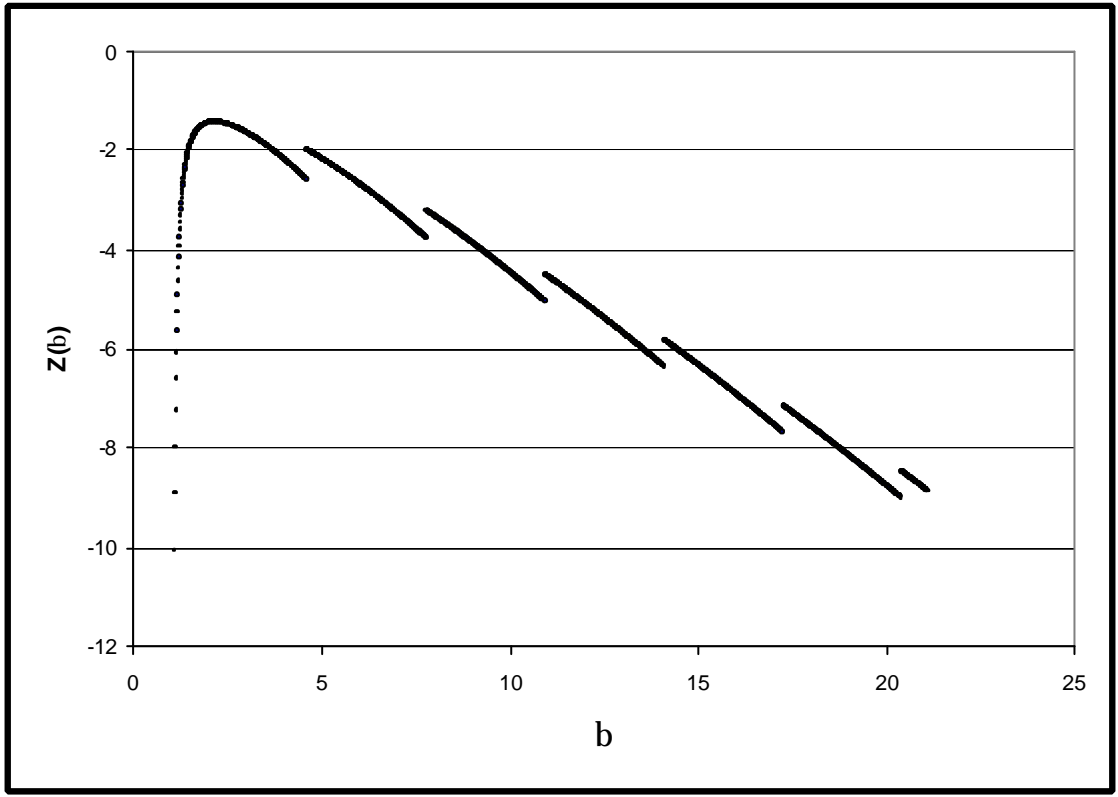

**Figure 3**  Coarsely sampled Plot of Z($\beta$) vs. $\beta$ from 0 to 21.  The discontinuities occur at the singularities in Table I.  Finer sampling reveals complex structure in a neighborhood of these singular $\beta$ values as illustrated in Figure 4.

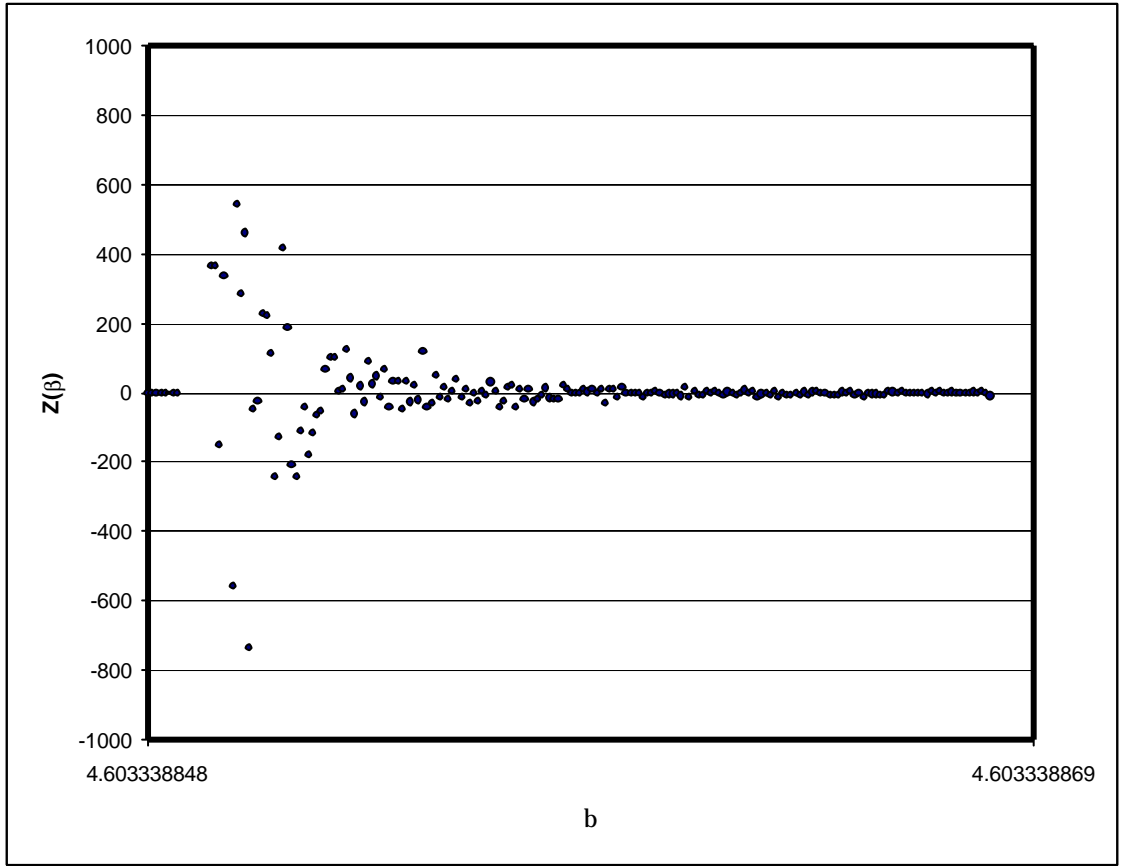

**Figure 4** A finely sampled Plot of Z($\beta$) vs. $\beta$ in the neighborhood of the first singularity showing positve values of Z corresponding to a net attractive radial force for some values of $\beta$.

## 7. The Causal Form for Electrodynamics

When considering the causal form for electrodynamics, the only change from the Feynman-Wheeler case is that the retarded potentials must be used in the field expressions instead of the average between the retarded and advanced potentials. We consider a particle which is constrained to move in a helix, and we calculated the electromagnetic force on it. It can be shown that the radial force is identical to the Feynman-Wheeler case but the azimuthal force is now nonzero. It is convenient to express the azimuthal force in terms of a ratio with the radial force. Figures 5 and 6 show plots of the calculated azimuthal forces for two ranges of $\beta$. These plots show that the sign of the ratio is always positive. When the radial force is attractive in sign, the azimuthal force is opposite in direction to the particle's velocity. This apparently represents radiative reactive force due to energy lost to radiation. It should be noted that it is well established [6,13] that as a tachyon loses kinetic energy, its speed increases.

$$F_{aximuthal} = F_r \mathrm{e}(\text{ß}) \tag{40}$$

We see from Figure 5 that as the velocity increases, the azimuthal force becomes smaller as compared to the radial force.

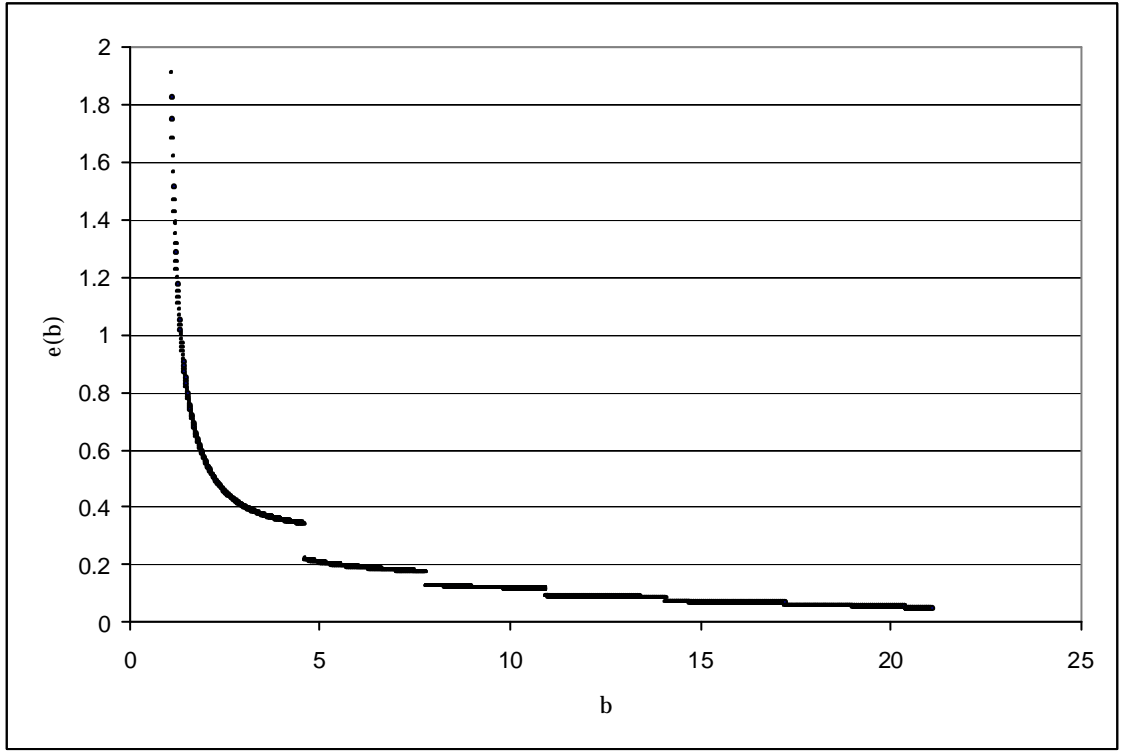

**Figure 5**  The ratio of azimuthal to radial force with causal (retarded) electrodynamics. The radial force is the same as for Feynman-Wheeler electrodynamics.

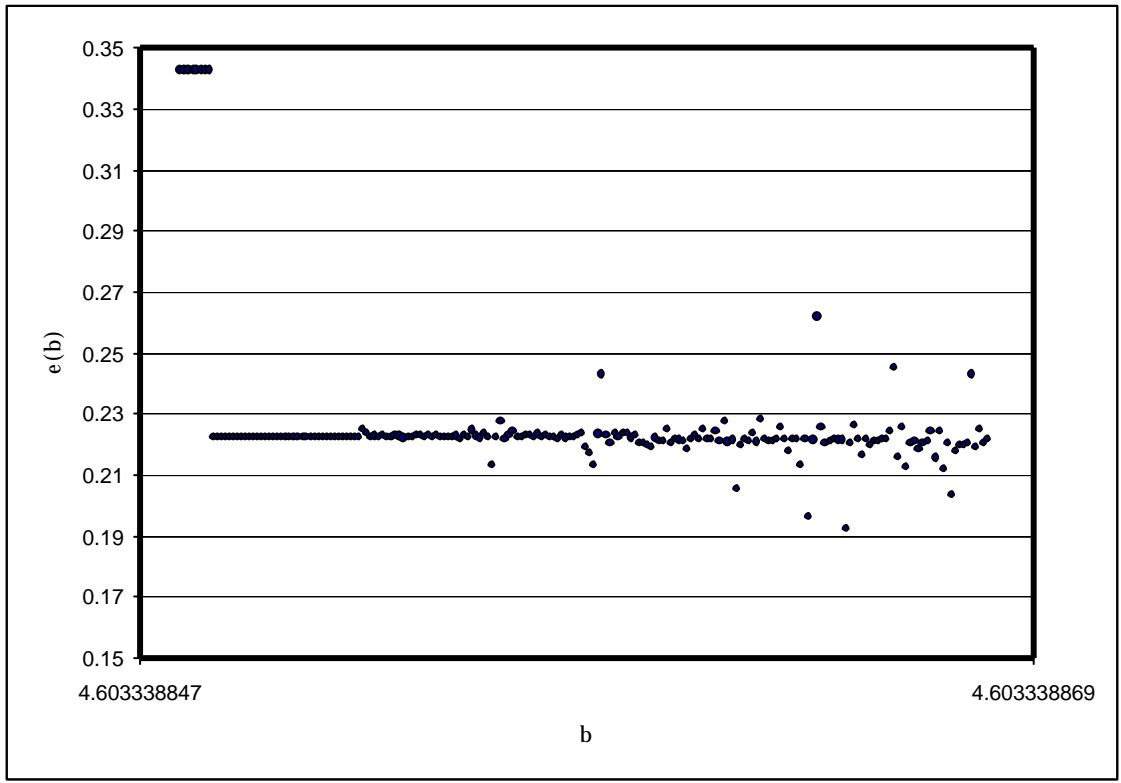

**Figure 6** Finely sampled ratio of Azimuthal to Radial force with Causal (retarded) Electrodynamics

## 8. Conjectural Implications

Theoretical conjectures are presented in the next few subsections. They present ideas which are of a hypothetical nature that serves to highlight the potential of this subject.

## 8.1 Radiation and radiative reactive forces

The results of the previous section show that when a tachyon crosses its own ligthcone it can experience a radiative force which causes its energy to change. This case was not considered by Ey and Hurst [5] who showed that tachyon's do not experience the same local radiative self-force as do bradyons in the Lorentz-Dirac equation. Ey and Hurst concluded from this that tachyons can never radiate. However, the above result shows that this conclusion must be amended. Tachyons can still radiate when they cross their own light cone. The Ey and Hurst result is still extremely important because it shows that the classical equations of motion for tachyons do not have the radiative reaction term of the Lorentz-Dirac equation. As a consequence, the Tachyon equation of motion does not appear to have the same problems with runaway solutions as the Lorentz-Dirac equation for bradyons. It's interesting to note that the runaway solutions of the Lorentz-Dirac equation involve cosh and sinh functions of the proper time for the time and one spatial coordinate [17]. This is mathematically equivalent to a circular motion which has been analytically continued to imaginary values of proper time and where one space coordinate has become timelike as a result of the superluminal transformation. Tachyons may be considered to have imaginary values of proper time, and so perhaps there is a geometrical relationship between the runaway solutions of the Lorentz-Dirac equation and the helical motions that we have found here. Perhaps the runaway solutions are actually required for extended lorentz invariance where superluminal boosts are allowed.[6,7]. Perhaps they are just the helical motions that we are describing here, but viewed from a superluminal frame in which the tachyon is moving slower than light.

## 8.2 Causality

Tachyons, if they exist, may always or almost always be bound up in helical motions similar to the solutions exhibited here in other confined orbiting motions. Then they would appear to be moving slower than light and to have intrinsic angular momentum. If this were the case, then the tachyons couldn't be used to send faster-than-light signals, and so the causality paradox would be simply resolved.

## 8.3 Relevance to hidden variable theories and stochastic models of quantum mechanics

The results here are applicable in the hidden variable effort of quantum mechanics. The hope that hidden variables may be the way of nature continues to attract prestigious adherents such as Gerard t'Hooft [18]. When one surveys the field though, one finds that despite great effort, nobody has really discovered a satisfactory derivation of quantum mechanics from a hidden variable perspective.

First of all, the Bell theorem [19-21] has stymied investigators because a hidden variable model must be able to explain nonlocality ie. superluminal connections. Bell's theorem is usually thought to be the most decisive objection to hidden variable models. Tachyons certainly provide for the possibility of nonlocality. But then the tachyon theory must somehow end up being causal. Tachyons bound up in helical motions (or other more complex confined motions) will be effectively causal.

It is a long-standing problem to include spin into hidden variable models, particularly in a relativistically covariant way. Again, the helical motion of tachyons is a natural candidate for spin. In fact, Recami and Salesi [22] have recently suggested that classical particles with spin may be the key ingredient in finding an explanation of the quantum mechanical potential as it appears in Bohm's hidden variable theory [23 and references therein]. The current theory is fully relativistically covariant, and an additional benefit is that there are apparently no runaway solutions plaguing this theory as in the Lorentz-Dirac equation.

With the Feynman-Wheeler interaction form the tachyon doesn't radiate when it moves in a helix, but with the causal interaction form energy is lost by the tachyon to radiation if the force is attractive. If the tachyon has more complicated confined solutions, it's possible that radiationless states may someday be found. It's very likely that if more complex confined solutions exist for the tachyons then they will exhibit chaos and this could be the origin of quantum indeterminism.

A quantum mechanical wave might naturally be visualized as the confined motion of a tachyon which moves ergodically in a short time to fill out the wave. This confined motion would be a superposition of helical motion with chaotic motion of the center of the helix moving rapidly around the wave's region of support. The wave function could collapse almost instantaneously in this picture because of the tachyon's great speed. Such chaotic motion might provide the stochastic behavior in hidden variable models like stochastic quantum mechanics (Nelson[24], Davidson[ 25-26], and references therein) . Indeed, tachyons are much better suited for a description of stochastic models of this type than are bradyons because of the singular nature of the Markov process used in these models and their singular velocities. A single tachyon acts in some ways like a many body problem because of its ability to interact with its own past positions at possibly a large number of points. This is a natural explanation of wave-particle duality. In a two slit diffraction experiment for example, the tachyon would have the opportunity of going through both slits many times. This avoids the usual arguments against the particle's ability to "know whether or not the other slit was open" when it passes through one of the slits.

Another approach to explore in looking for a hidden variable model would be to postulate the classical zero-point vacuum radiation model of stochastic electrodynamics [27,28] and then analyze how the tachyon would diffuse in this background. The motion would be random due to the interaction with the background radiation, and also could be chaotic from its own self-interaction.

The Aharanov-Bohm effect [29] is another perplexing phenomenon that needs to be explained by hidden variable models. In the analysis above it was noted that singularities occur in the field expressions when the Cerenkov cone from a retarded source charge points to the test point. These singularities have no analog for slower than light particles. If the electrons in a solenoid were actually tachyons moving in helices, then they would produce a great multitude of singular field points even outside the solenoid. These might be hard to detect in most situations, as they might be randomly oriented and quickly changing with time and in most cases they might add up to zero net force on moving particles outside the solenoid. But maybe the Aharanov-Bohm experiment is a case where the effect can be measured. The point is that the Cerenkov singularity is a mechanism for something real to penetrate the region of the solenoid and affect a particle outside of it. Perhaps it can explain the Aharanov-Bohm effect.

The helical solutions that we have found may behave when perturbed in some ways like relativistic strings when their velocity tends to infinity. A circular motion will map out a helix in space time and if the velocity of the tachyon is very high, then the pitch of the helix will be small, and the tachyon's worldline will look like a cylindrical surface. Its mechanical behavior may begin to resemble that of a relativistic string in this limit.

Another problem for a hidden variable theory is how one is to interpret the double-valued nature of the half integral spin variable. Perhaps this too has an interpretation in the present theory. If one considers superluminal boosts added to the Lorentz group as is done in the extended relativity [6,7], then maybe it is possible to rotate through 360 degrees by doing it in the superluminally boosted frame, and then come back to the original unboosted frame. In some circumstances the classical coordinates may not return to their original values in this case. The reason is that the expression for proper time has a square root singularity in the velocity variable. Superluminal boosts may involve an analytic continuation in the proper time variable around a square root singularity which could yield the double-valued nature of spin in a classical albeit tachyonic framework. The double-valued nature of half-integral spin might be a result of the double-valued nature of the square root singularity for proper time when analytically continuing it to superluminal frames. It's plausible that when the tachyon circles around the axis of spin, then the spin is double-valued but when the helix itself circles slowly around an axis, it generates a single-valued angular momentum analogous to orbital angular momentum in quantum mechanics. We defer consideration of this idea to a future publication.

So we have a possible way around several of the most disconcerting objections that have been raised against hidden variable theories. Even it some of these conjectures prove incorrect, they can serve as a guide for those who are seeking a hidden variable description of quantum mechanics. If one is so inclined, it's extremely difficult to resolve even one of the above paradoxes in a classical theory. The present theory potentially has a resolution for all of them. This is truly remarkable and unique. Moreover, from the point of view of Occam's razor, the present tachyon theory is very economical. It posits only pointlike charged particles. What other explanation for the existence of spin could be this simple in a hidden variable theory based on classical physics?

Obviously, helical motions may not be the only confined motions for tachyons in this theory, but they are probably the simplest. The analysis of section 5 shows that when the tachyon's past trajectory is nearly tangent to the past light cone as drawn from the tachyon's current position, then the resultant force is the difference between two large numbers and is thus very sensitive to the current position of the tachyon. An example of this hyper-sensitivity is shown in Figure 4. The motion of the tachyon if its orbit is perturbed from a perfect helix at nearly a critical velocity can be expected to be complicated because of this force, and in some situations my be effectively nondeterministic. The tachyon must overcome an infinite barrier to reach a position where its past trajectory is exactly tangent to the past light cone. This fact may prevent the tachyon's orbit from ever unraveling.

It is very likely that when one analyzes two or more tachyons, that circular solutions to their motions will be found in which the group of tachyons move together in a circle.

## 9. Conclusion

This theory may have implications for hidden variable theories of quantum mechanics as many paradoxes or difficulties are conceptually overcome with tachyons in helical orbits as models for elementary particles. Charged tachyons may be masquerading as slower than light particles because they are moving in tight helices which disguise their faster-than-light motion. These particles would appear to have intrinsic angular momentum (spin) and magnetic moments, even though the actual charged tachyon had no spin.

If a tachyon moves in a helix, then the center of the helix will transform as a slower than light particle under Lorentz transformations, and therefore it will always move forward in time on the average no matter what frame it is viewed in. This could largely mitigate the causality problem for tachyons whereby they could be used to send signals backward in time as in the Tolman Paradox [8]. If they move in tight helical motion, they can't easily be used to signal backwards in time.

## Acknowledgements

The author acknowledges extensive discussions with Mario Rabinowitz, useful conversations with Vladimir Kresin, and helpful correspondence with Erasmo Recami, Angas Hurst, and Ed Nelson.

## Figure Captions

**Figure 1** A tachyon moving in a helix and intersecting its own light cone

**Figure 2** Light cone intersection points come in symmetrical pairs, one point the intersection with the retarded light cone and the other the intersection point of the advanced light cone.

**Figure 3** Coarsely sampled Plot of $Z(\beta)$ vs. $\beta$ from 0 to 21. The discontinuities occur at the singularities in Table I. Finer sampling reveals complex structure in a neighborhood of these singular $\beta$ values as illustrated in Figure 4.

**Figure 4** A finely sampled Plot of $Z(\beta)$ vs. $\beta$ in the neighborhood of the first singularity showing positve values of Z corresponding to a net attractive radial force for some values of $\beta$.

**Figure 5** The ratio of azimuthal to radial force with causal (retarded) electrodynamics. The radial force is the same as for Feynman-Wheeler electrodynamics.

**Figure 6** Finely sampled ratio of Azimuthal to Radial force with Causal (retarded) Electrodynamics

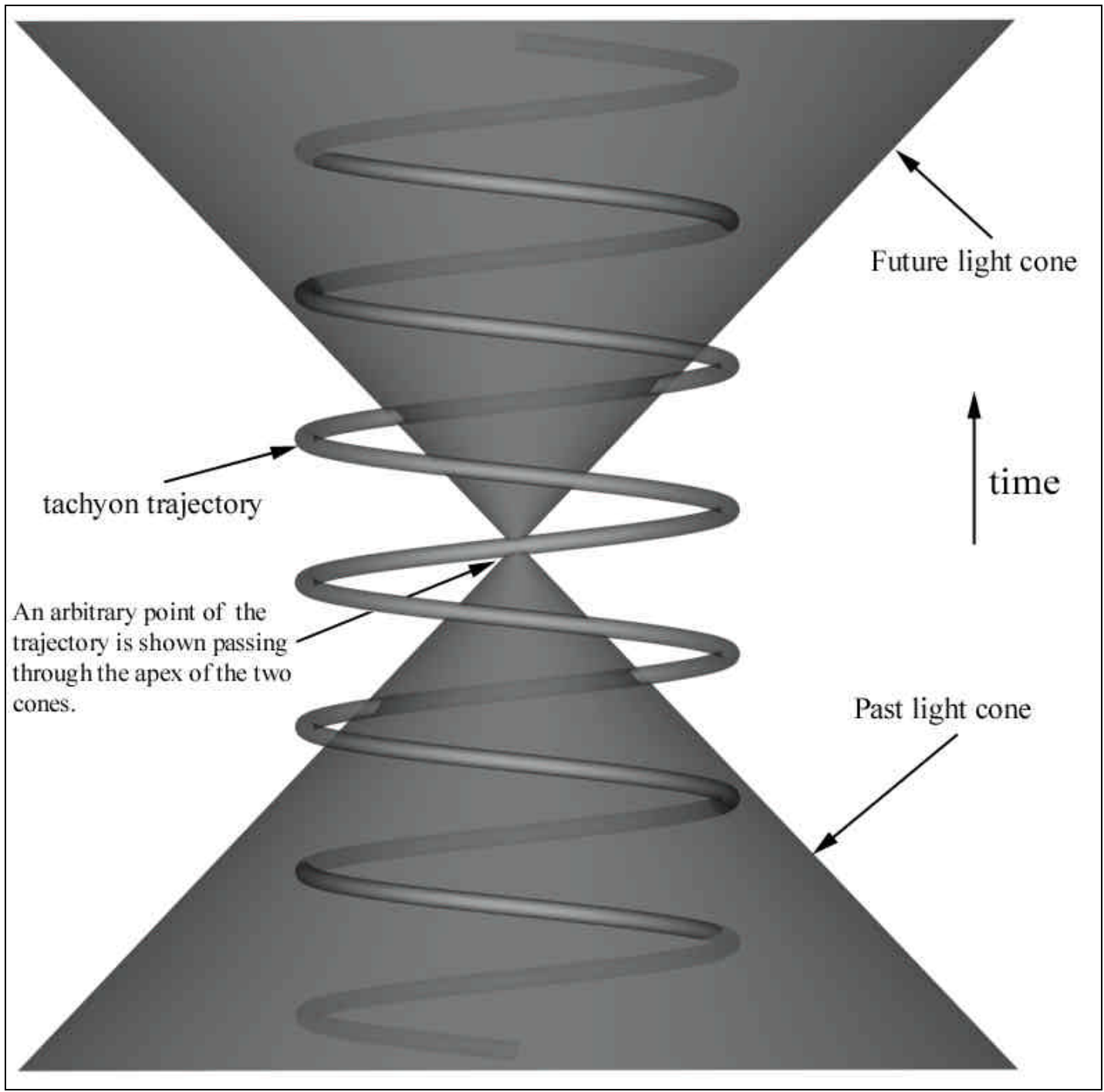

Future light cone

time

tachyon trajectory

An arbitrary point of the
trajectory is shown passing
through the apex of the two
cones.

Past light cone

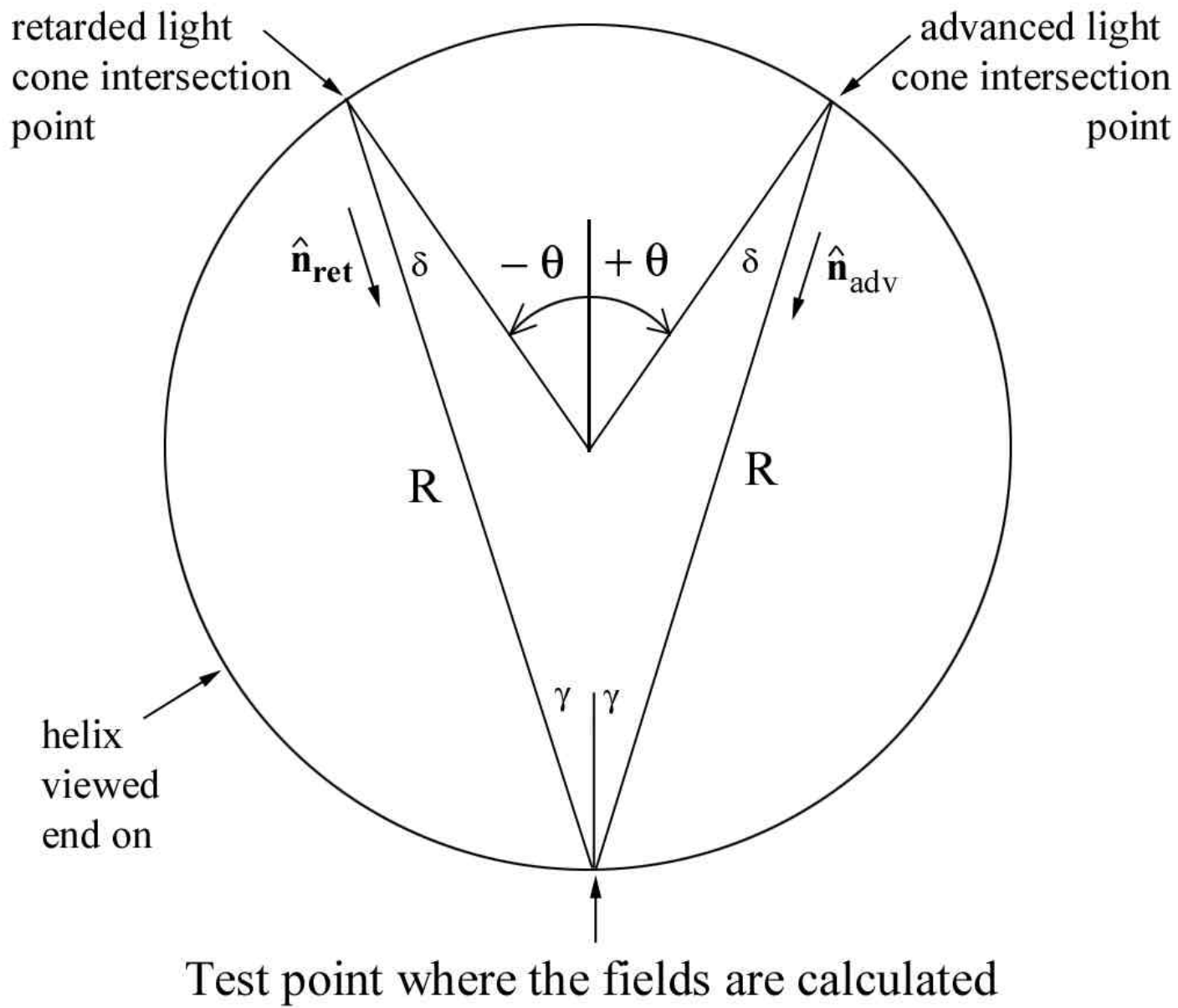

retarded light
cone intersection
point

advanced light
cone intersection
point

$\hat{\mathbf{n}}_{\mathbf{ret}}$   $\delta$   $-\theta$   $+\theta$   $\delta$   $\hat{\mathbf{n}}_{\mathrm{adv}}$

$R$       $R$

$\gamma$   $\gamma$

helix
viewed
end on

Test point where the fields are calculated

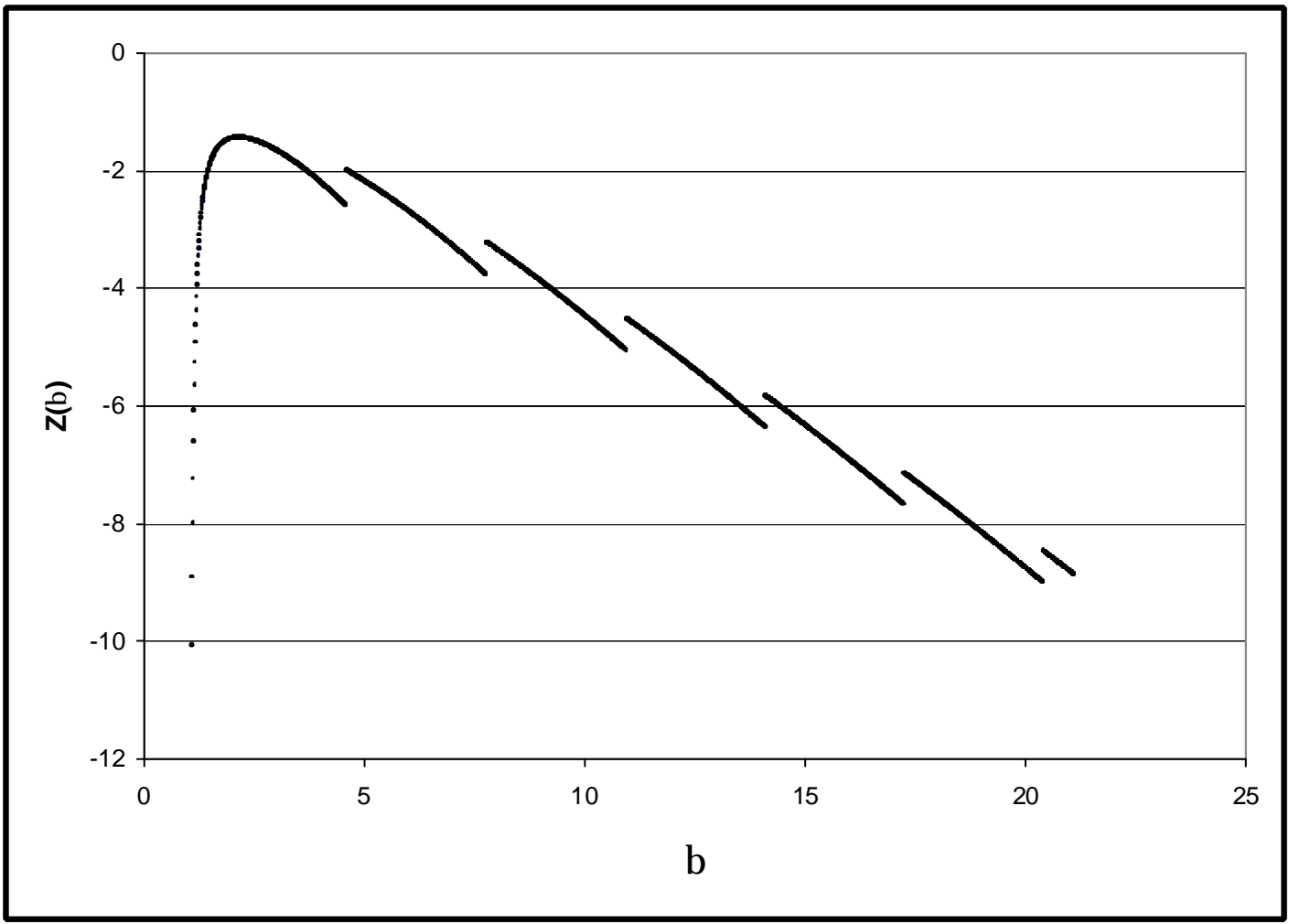

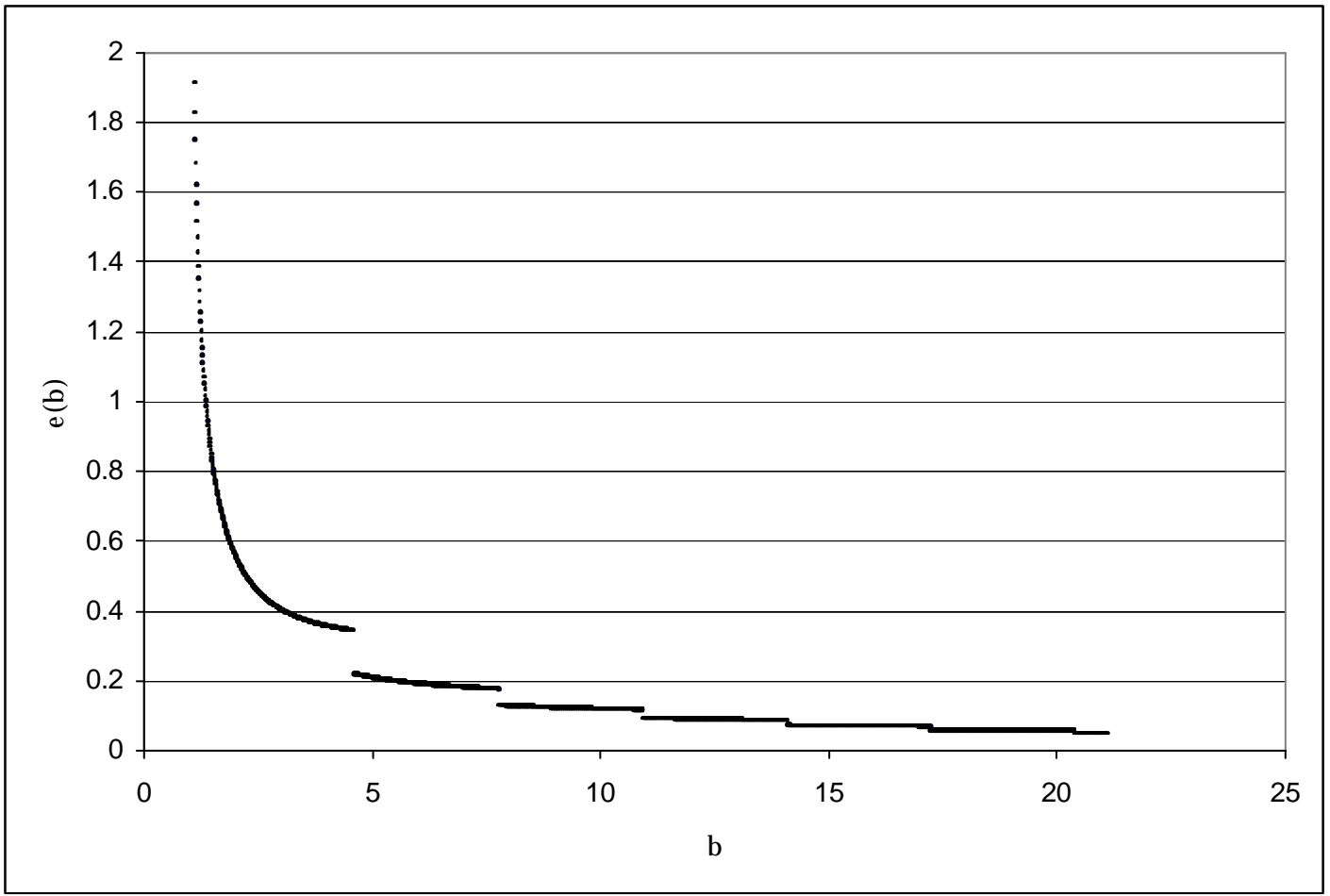